\documentclass[a4paper,11pt]{article}
\usepackage{pos}

\title{Evaluation of $SU(3)$ smearing on FPGA accelerator cards}

\author[a]{Salvatore Cal\`i}
\author*[b]{Grzegorz Korcyl}
\author[c]{Piotr Korcyl}

\affiliation[a]{Center for Theoretical Physics, Massachusetts Institute of Technology, Cambridge, MA 02139, USA}

\affiliation[b]{Institute of Applied Computer Science, Jagiellonian University, ul.
prof.  {\L}ojasiewicza 11, 30-348 Krak{\'o}w, Poland}

\affiliation[c]{Institute of Theoretical Physics, Jagiellonian University, ul.
prof.  {\L}ojasiewicza 11, 30-348 Krak{\'o}w, Poland}

\emailAdd{calis@mit.edu}
\emailAdd{grzegorz.korcyl@uj.edu.pl}
\emailAdd{piotr.korcyl@uj.edu.pl}

\abstract{Recent FPGA accelerator cards promise large acceleration factors for some specific computational tasks. In the context of Lattice QCD calculations, we investigate the possible gain of moving the $SU(3)$ gauge field smearing routine to such accelerators. We study Xilinx Alveo U280 cards and use the associated Vitis high-level synthesis framework. We discuss the possible pros and cons of such a solution based on the gathered benchmarks.\\
\flushright{MIT-CTP/5341} 
}

\FullConference{%
 The 38th International Symposium on Lattice Field Theory, LATTICE2021
  26th-30th July, 2021
  Zoom/Gather@Massachusetts Institute of Technology
}

\begin{document}
\maketitle

\section{Introduction}


As the computer architectures become more and more heterogeneous it may be advantageous to delegate some steps of the calculations to different resources present on the cluster/supercomputer nodes. In such scenario some elements could be executed in parallel by different architectures. For instance, in CYGNUS installation, preprocessing of data for data exchanges is accelerated by the FPGA processors. With this in mind, in the context of lattice QCD, we benchmark the APE link smearing routine on the Xilinx Alveo U280 accelerator card. 

\begin{figure}[b]
    \centering
    \includegraphics[width=0.2\textwidth]{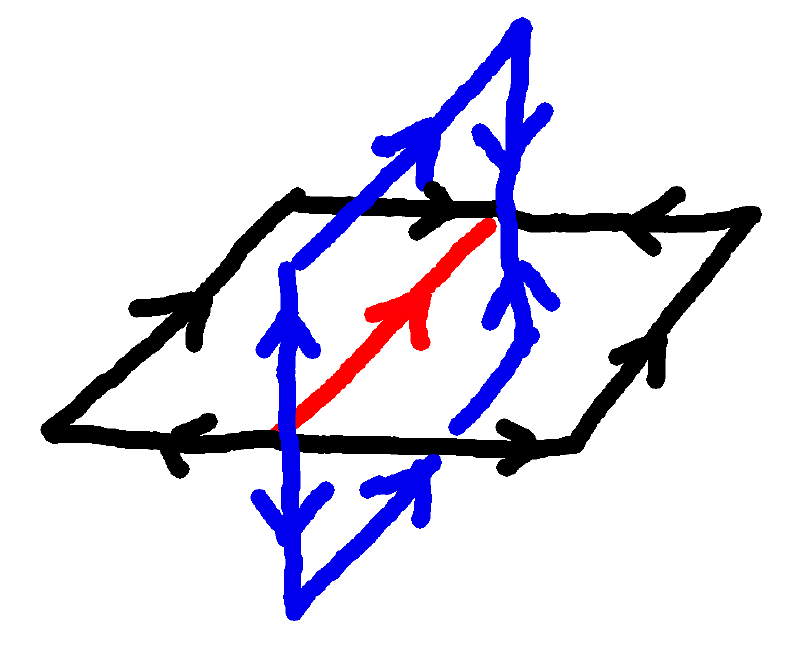}
    \caption{Schematic representation of the APE link smearing. The link is a $SU(3)$ matrix and is a basic degree of freedom. The link being smeared is marked in red. Two blue and two black "staples" are shown, each one being a product of three links. The full smearing routine contains another pair of "staples" in the fourth direction.}
    \label{fig: ape smearing}
\end{figure}

The APE smearing \cite{APE:1987ehd} is a representative case of input data averaging defined by a 9-point stencil on a data grid with a topology of a four dimensional torus. In lattice QCD the basic degrees of freedom located on the edges of the grid are $3\times3$ complex values matrices belonging to the $SU(3)$ group, called "links". Because of the non-abelian nature of that group, averaging of neighbouring parallel links is replaced by the average of "staples", i.e. products of three link variables along the lines sketched in Figure \ref{fig: ape smearing}. For each link one needs to evaluate 6 staples and perform a substitution,
\begin{equation}
    U_{\mu}(x) \rightarrow U_{\mu}(x) + \sum_{i=-3}^3 S_i(x) 
    \label{eq: smearing}
\end{equation}
where $S_{\pm 1}$,$S_{\pm 2}$ and $S_{\pm 3}$ are the staples in three directions perpendicular to the direction of the link $U_{\mu}(x)$. $\pm$ corresponds to the two possibilities: "up" or "down", "left" or "right" which we denote in the following altogether by "forward" and "backward". Eq.~\eqref{eq: smearing} differs from the common definition in the Literature by scaling coefficients which all were set to 1. Such coefficients are irrelevant as far as performance is concerned.

From the point of view of a compute node, we assume that the host CPU supervises the main compute flow and delegates parts of the computations to different devices. Hence, we assume that the gauge links have been transferred from the host to the High Bandwidth Memory (HBM) memory of the FPGA accelerator. The described implementation takes the input link variables which are streamed to the programmable logic from the HBM, transforms them and stores back in the HBM memory. This process can be iterated. Ultimately, the smeared link variables are transferred back to the host. Below we describe the details of the FPGA kernel and data transfer mechanisms. Our work is built on previous implementations of the CG solver \cite{Korcyl:2018pjc, Korcyl:2018vad, Korcyl:2019uli, Korcyl:2020veo}. For recent progress in the FPGA optimized HPCG benchmark see Ref.~\cite{zeni2021optimized}. 

\section{Pipelined and streamlined design}

\begin{figure}[ht]
\begin{center}
    \includegraphics[width=0.4\textwidth]{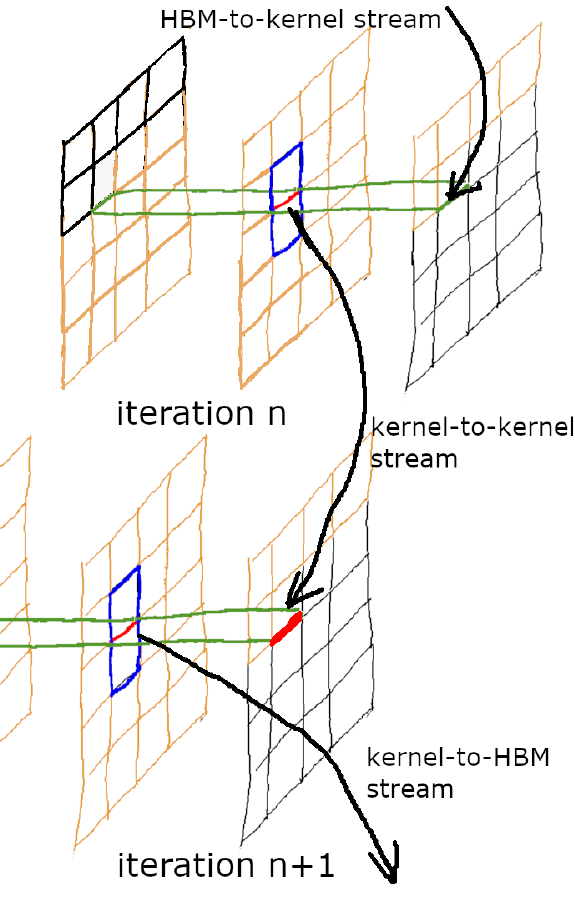}
\end{center}
\caption{Schematic view of data flow in the HBM-kernel-kernel-HBM stream with cyclic buffers (orange) implemented in the U/BRAM}
\label{fig: iteration pipeline}
\end{figure}

In order to fully exploit the possibilities given by the U280 accelerator one has to consider and implement several levels of parallelism. At the lowest level, we have data parallelism which we can realize by instantiating several instances of a kernel to process multiple data simultaneously. For instance, staples in three directions can be evaluated in parallel if we instantiate three separate kernels calculating staples (see Table \ref{tab: resources1}). At one-step higher level, one can exploit parallelism in time by pipelining the computations. Again, let us take a computation of a single staple as example. Its evaluation in double precision takes 39 clock cycles (again, see Table \ref{tab: resources1}). By using special directives from the Vitis environment, we can instruct the compiler to produce a kernel which can be fed with new data every Initiation Interval (II) clock cycles (see fourth column of Table \ref{tab: resources1}). In the case of double precision this can be $\textrm{II}=2$. This means that, at a given moment of time, the kernel responsible for the staple evaluation will be performing computations for $39/2 \approx 20$ staples in parallel. Eventually, since typically the smearing algorithm involves many iterations of the same procedure on the same data, one can construct a pipelined data flow using multiple instances of the entire smearing routine kernel in such a way that in a given moment of time multiple iterations will be executed in the FPGA accelerator. This latter idea is schematically depicted in Figure \ref{fig: iteration pipeline}. The plot shows slices of the lattice with the link being smeared marked in red. The necessary staples are shown in blue and green. The upper part represents one kernel implementing one iteration of the smearing routine; the lower part is a second, separate kernel implementing the second iteration. Data flow is marked with black arrows: original data arrives in a stream from the HBM to the programmable logic, it is processed by the first kernel performing the iteration $n$, subsequently it is sent in a form of another stream to the second kernel where the iteration $n+1$ is executed. Finally, the data is streamed back to the HBM memory.
The link variables shown in orange on the sketch are kept in the local memory of the kernel in an array in the form of a FIFO cyclic buffer. The black link variables have already been used and were removed from the buffer, the grey will be transferred to the kernel in the next steps of the volume loop. Although we have implemented and tested this mechanism, we did not manage to compile the entire project including the cyclic buffers with all the constraints, because of local congestion problems in the HBM-Super Logic Region (SLR) region. Hence, although the U280 has enough resources to implement the entire project, the performances quoted in the following section are based on partial compilation results.

Combining all three levels of parallelism together with the corresponding data transport layers allows to fully exploit the potential of the FPGA accelerators. In practice, the feasibility of the project depends on: the size of the available resources which we discuss in the next section and on the ability of the compiler to efficiently implement everything within the time and space constraints, on which we comment in the last section.

\section{Resource consumption}

\begin{table}[ht]
\begin{center}
\caption{Composition of multiply\_by\_staple function}
\begin{tabular}{|ccccccc|}
\hline
component & \# & latency & interval & DSP & FF & LUT \\
\hline
grp\_compute\_staple\_forward\_fu	    & 3 & 39    & 2 & 400	& 31231	& 21159	 \\
grp\_compute\_staple\_backward\_fu	& 3 & 39    & 2	& 400	& 31231	& 21159	\\
grp\_add\_two\_fu	                    & 3 & 4	    & 1	& 36	& 2827	& 2178 \\
\hline
\end{tabular}
\label{tab: resources1}
\end{center}
\end{table}

\begin{table}[ht]
\begin{center}
\caption{Comparison of the resource consumption of various compute kernels for different data types, compiled for U280 card at 300 MHz with Vitis HLS 2020.2 (resources in \% of total / of one SLR)}
\begin{tabular}{|cccccccc|}
\hline
function & prec. & latency & II & BRAM & DSP &	FF &	LUT \\
\hline
compute\_staple\_forward              & \textbf{double} & 65 & \textbf{2} & 0   & 12 / 37    & 5 / 16	& 6 / 18 \\
compute\_staple\_forward              & \textbf{double} & 67 & \textbf{4} & 0   & 6 / 18    & 3 / 10	& 3 / 9 \\
compute\_staple\_forward              & \textbf{double} & 71 & \textbf{8} & 0   & 3 / 9    & 2 / 6	& 1 / 5 \\
multiply\_by\_staple         & \textbf{double} & 90 & \textbf{2} & 0   & 77 / 231	& 34 / 103	& 39 / 118 \\
multiply\_by\_staple         & \textbf{double} & 93 & \textbf{4} & 0   & 38 / 115	& 21 / 63	& 21 / 63 \\
multiply\_by\_staple         & \textbf{double} & 99 & \textbf{8} & 0   & 19 / 57	& 13 / 41	& 11 / 34 \\
\hline
compute\_staple\_forward              & \textbf{float} & 69 & \textbf{2} & 0    & 5 / 16    & 2 / 7	& 2 / 7	\\
compute\_staple\_forward              & \textbf{float} & 72 & \textbf{4} & 0    & 2 / 8    & 1 / 4	& 1 / 4	\\
multiply\_by\_staple         & \textbf{float} & 100 & \textbf{2} & 0    & 17 / 104	& 15 / 47	& 16 / 50 \\
multiply\_by\_staple         & \textbf{float} & 105 & \textbf{4} & 0    & 17 / 52	&  10 / 30	& 9 / 29 \\
\hline
compute\_staple\_forward              & \textbf{half} & 72 & \textbf{2} & 0	  & 4 / 13	    & 1 / 4	& 1 / 4 \\
multiply\_by\_staple         & \textbf{half} & 103 & \textbf{2} & 0     & 27 / 82	& 10 / 31	& 9 / 29 \\
\hline
\hline
su3\_projection                      & \textbf{double} & 869 & \textbf{8} & 0   & 14 / 43	& 10 / 31	& 8 / 26 \\
su3\_projection                      & \textbf{float} & 899 & \textbf{4} & 0    & 13 / 39	& 7 / 23	& 7 / 23 \\
su3\_projection                      & \textbf{half} & 909 & \textbf{2} & 0     & 20 / 62	& 8 / 24	& 7 / 23 \\
\hline
\hline
full                      & \textbf{double} & 989 & \textbf{8} & 0   & 33 / 100	& 25 / 75	& 20 / 62 \\
full                      & \textbf{float} & 1022 & \textbf{4} & 0    & 30 / 91	& 17 / 53	& 17 / 53 \\
full                      & \textbf{half} & 1037 & \textbf{4} & 0     & 24 / 73	& 11 / 35	& 11 / 35 \\
full                      & \textbf{half} & 1014 & \textbf{2} & 0     & 49 / 147	& 19 / 57	& 17 / 53 \\
\hline
\end{tabular}
\label{tab: resources2}
\end{center}
\end{table}

The feasibility of the implementation outlined in the previous section depends on the size (in terms of logical elements resources) of the single kernel. In our implementation the kernel is composed of several modules instantiated as separate functions: $SU(3)$ group elements scaling by a scalar, addition (\verb[add_two[) and multiplication, evaluation of a single staple (\verb[compute_staple_*[), multiplication by the sum of staples (\verb[multiply_by_staple[) and projection back to the $SU(3)$ group (\verb[su3_projection[). On one hand, the best performance is obtained when all the functions are merged by the \verb[inline[ keyword allowing for the compiler to reshuffle and reuse resources and avoid constructing interfaces for consecutive functions calls. On the other hand, when each function is left as a separate module, the compiler provides individual information on resources consumption which allows to understand which elements are critical from the point of view of resource consumption and also which functions are reusing the same instances of lower-level kernels. Following the second possibility, we gather relevant information on the resource consumption of the various steps of the smearing procedure in Table \ref{tab: resources1} and \ref{tab: resources2}. In order to estimate the total performance we use inlining for all functions.

As an example, Table \ref{tab: resources1} shows the structure of the \verb[multiply_by_staple[ function which yields the product of the current link and the sum of the six staples at a one level decomposition. We see that the compiler has generated three instances of the kernels \verb[grp_compute_staple_forward_fu[, \verb[grp_compute_staple_backward_fu[ and \verb[grp_add_two_fu[, which already signifies that the evaluation of the six staples will be performed in parallel. The inner structure of these functions is hidden at this point, but may be unraveled if we unset the \verb[inline[ keyword for them. In that case we would be able to monitor how the parallelism of the $SU(3)$ matrix multiplications is implemented in the logic. Vitis software allows to control the number of instances of each function and hence the user can directly reduce/increase the resource consumption to reduce/increase the parallelism. The fourth column of Table \ref{tab: resources1} contains data on the initiation interval which is directly proportional to the total performance.

In Table \ref{tab: resources2} we show the resource consumption and the latency and initiation interval of all the higher-level functions from the smearing routine as a function of the data precision (column 2) and imposed initiation interval (column 4), both highlighted with bold letters. The initiation interval can be controlled from the Vitis environment by a special pragma. The smaller is the II, the larger the performance. At the compilation stage, although the compiler can produce a kernel with a given II, we may not be able to provide input data at that speed or the resources needed to sufficiently parallelize the kernel to keep up to this II may not be available. The latter turns out to be the case for the kernel \verb[multiply_by_staple[ in double precision with II=2 which exceeds the DSP, FF and LUT resources in a single SLR. With $\textrm{II}=4$ the number of needed DSP is exceeded, which also rules out this setup. Similar observations may be done for the same kernel in single precision with $\textrm{II}=2$. From that point of view, we conclude that the possible II for double precision is $\textrm{II}=8$, for float is $\textrm{II}=4$ and for half is $\textrm{II}=2$. This conclusion will be confirmed by the analysis of the input data bandwidth which we discuss in the next section. The full size of the smearing routine, composed of the staple evaluation and multiplication and of the $SU(3)$ projection, is shown in the last four rows of Table \ref{tab: resources2} only for the parameters which fit in a single SLR.

\section{Timings and performance}

In order to assess the performance of the setup presented above one has to count the number of floating point operations needed for the smearing of a single link. The input data is composed of six sets of three $SU(3)$ matrices needed for the six staples. Hence, for each link we need to load $18 \times 9 \times 2 = 324$ floating point numbers. For each staple we have two matrix-matrix multiplications, hence 12 multiplications and 6 matrix-matrix additions. This gives $324 \times 12+108=3996$ floating point operations (FLOPs). Finally, the $SU(3)$ projection \cite{projection} requires $2790$ FLOPs where the number of iterations was set to 4.

As far as the data transfer is concerned, the HBM memory on the Xilinx U280 card has 32 512-bit wide ports which can run at 300 MHz. The 32 ports are divided equally among four regions of the programmable logic (SLR). From the point of view of possible paths congestion it is advisable not to exceed one SLR and work with 8 ports attached to it. In Table \ref{tab: initiation interval} we provide the size of the input in bits for the different precisions. In the second column we translate the latter into the number of 512-bit words which have to be transferred. Finally, in the third column we report the minimal (when all 8 ports are used) and maximal (when only a single port is used) number of clock cycles needed to transfer input data for the smearing routine of a single link variable. This number of clock cycles directly translates into the initiation interval for the kernel, since we cannot start the kernel before all the data has arrived. The last column contains the final initiation interval for the given precision, chosen in accordance with the resource consumption presented in the previous section.

\begin{table}[ht]
    \centering
    \begin{tabular}{|ccc|c|c|}
    \hline
     & total size & \# 512-bit & initiation & optimal \\
     & in bits & words &   interval & II\\
    \hline
    double & 4608 & 9 & 2 - 9 & \textbf{8} \\
    float & 2304 & 4.5 & 2 - 5 & \textbf{4}\\
    half & 1652 & 2.25 & 2 - 3 & \textbf{2} \\
    \hline
    \end{tabular}
    \caption{Possible values of the initiation interval inferred from the HBM-programmable logic bandwidth.}
    \label{tab: initiation interval}
\end{table}
With the initiation interval fixed by the available resources and memory bandwidth we can estimate the performance of a single kernel. We have gathered the numbers in Table \ref{tab: performance}.
\begin{table}[ht]
    \begin{center}
    \begin{tabular}{|c|c|c|c||c||c|}
    \hline
    precision & II & staples & projection & full & 3 kernels\\
         & & [GFLOP/s] & [GFLOP/s] & [GFLOP/s] & [GFLOP/s] \\
    \hline
    double & 8 & 150 & 105 & 255 & 765 \\
    float & 4 &  300 & 210 & 510 & 1530 \\
    half &  2 & 600 & 420 & 1020 & 3060 \\
    \hline
    \end{tabular}
    \end{center}
    \caption{The initiation interval, inferred from the HBM-programmable logic memory bandwidth, sets the performance limit on a single kernel. The last column provides estimates in GFLOP/s, assuming that 3 parallel kernels are implemented in 3 separate SLR domains.}
    \label{tab: performance}
\end{table}

We can contrast these numbers with our benchmark runs performed on the Prometheus supercomputer hosted by the AGH Cyfronet in Kraków, Poland. Each node is equipped with a two-socket, 24-core Intel Haswell processor. 50 iterations of the APE smearing on a lattice of size $32^3 \times 64$ using 6 nodes took 3.0s, which translates into 110 GFLOPs/s per node.









\section{Conclusions and outlook}

In this work we have evaluated the performance of the APE smearing routine executed on the Xilinx Alveo U280 accelerator. Our implementation exploits several layers of parallelism offered by FPGA accelerators as well as the benefits of HBM memory located close to the programmable logic. Our analysis shows that a speedup factor compared with CPU is possible, provided the compilation, placement and routing of all elements is successful. 
Although we have tested all the elements individually and the SLR domain of Alveo U280 is large enough to contain the complete solution, we did not yet manage to obtain the final binary, due to Vitis 2020.2 failing in placing and routing the generated logic resources, because of high level of congestion. The problem remains still open and the solution will be evaluated with various Vitis releases, which highly differ in delivered quality of results.
Work in this direction is still being done. Also, as some additional research direction it would be interesting to benchmark the SyCL framework for FPGA with the code described here.

\section*{Acknowledgements}

This work was supported by the Foundation for Polish Science grant no. TEAM/2017-4/39, by the Polish Ministry for Science and Higher Education grant no. 7150/E-338/M/2018, and by the Priority Research Area Digiworld under the program Excellence Initiative – Research University at the Jagiellonian University. We gratefully acknowledge hardware donations from Xilinx within the Xilinx University Program. S.C. acknowledges support from the Carl G and Shirley Sontheimer Research Fund at MIT.

\bibliographystyle{acm}
\bibliography{references}



\end{document}